# Comparing Human Expertise and Large Language Models Embeddings in Content Validity Assessment of Personality Tests

*Nicola Milano*[1], *Michela Ponticorvo*[1], *Davide Marocco*[1]

[1]University of Naples Federico II

Department of Humanistic Studies

Natural and Artificial Cognition Laboratory "Orazio Miglino"

via Porta di Massa 1, Naples, 80125, Italy

Author Note

Correspondence concerning this article should be addressed to Nicola Milano.

Email: nicola.milano@unina.it

**Abstract**

In this article we explore the application of Large Language Models (LLMs) in assessing the content validity of psychometric instruments, focusing on the Big Five Questionnaire (BFQ) and Big Five Inventory (BFI). Content validity, a cornerstone of test construction, ensures that psychological measures adequately cover their intended constructs. Using both human expert evaluations and advanced LLMs, we compared the accuracy of semantic item-construct alignment. Graduate psychology students employed the Content Validity Ratio (CVR) to rate test items, forming the human baseline. In parallel, state-of-the-art LLMs, including multilingual and fine-tuned models, analyzed item embeddings to predict construct mappings. The results reveal distinct strengths and limitations of human and AI approaches. Human validators excelled in aligning the behaviorally rich BFQ items, while LLMs performed better with the linguistically concise BFI items. Training strategies significantly influenced LLM performance, with models tailored for lexical relationships outperforming general-purpose LLMs. Here we highlights the complementary potential of hybrid validation systems that integrate human expertise and AI precision. The findings underscore the transformative role of LLMs in psychological assessment, paving the way for scalable, objective, and robust test development methodologies.

1. Introduction

Recent advancements in artificial intelligence, particularly the development of transformer-based Large Language Models (LLMs) such as GPT and BERT (Vaswani et al., 2017; Radford et al. 2018, 2019; Brown et al., 2020; OpenAI, 2023), have revolutionized the field of natural language processing and introduced novel methodologies in psychological research. These models, trained on vast corpora of text, create high-dimensional semantic embeddings that capture the contextual and linguistic nuances of sentences rather than individual words. Such capabilities have paved the way for innovative applications in psychometrics, including the semantic analysis of test items to predict and validate latent factorial structures in personality assessments. For instance, recent studies have demonstrated how LLMs can effectively cluster semantically similar items under their hypothesized constructs, achieving alignment with human-derived factorial structures without relying on response data (Wulff and Mata, 2024; Guenole et al, 2024, Milano et al., 2025). The focus on the connection between linguistic items and psychological constructs can be useful in research on validity as well.

In psychometric research, indeed, validity is a core theme and it represents the degree to which a test accurately measures what it purports to measure, and it is a fundamental aspect of ensuring that psychological assessments yield meaningful, results. Validity is paramount for the utility of any psychometric tool, as it directly influences the interpretation of test scores and the decisions that follow (Messick, 1995). The consequences of inadequate validity are particularly significant in high-stakes assessments, such as intelligence testing, educational evaluations, or clinical diagnostics. Erroneous assessments can lead to misclassification, inappropriate interventions, or inaccurate data for research purposes (Anastasi & Urbina, 1997). Validity, therefore, ensures that tests provide not just consistent results (reliability) but correct results that align with the theoretical and practical applications of the measurement tool (American Educational Research Association et al., 2014).

In summary, validity is indispensable to psychometrics because it underpins the trustworthiness and scientific integrity of the inferences drawn from psychological assessments. Without valid tests, the field would lack the precision needed to advance theory, inform evidence-based practice, and guide effective interventions.

Moving from a general description of validity, to a more specific one, it is useful to underline that it is a multifaceted concept, often subdivided into several types—construct validity, content validity, and criterion-related validity, among others.

In this paper, we will focus on content validity, that assesses whether the test adequately covers the domain of interest and on how LLMs can help to answer content validity-related questions. Content validity reflects the extent to which a measurement tool adequately covers the entire domain of the construct it intends to assess and it is therefore linked on how items cover the construct.

Unlike other types of validity, which often rely on statistical analyses, content validity is evaluated through expert judgment, focusing on the representativeness and comprehensiveness of test items within the defined construct (Haynes, Richard, & Kubany, 1995). In practice, establishing content validity involves defining the construct thoroughly and ensuring that all relevant facets are reflected in the test items. For instance, a test designed to measure mathematical aptitude must encompass a range of items that cover different areas within mathematics, such as algebra, geometry, and arithmetic, rather than disproportionately focusing on one area (Sireci, 1998). A content-valid test should also avoid including extraneous items that do not align with the construct, as these can reduce the test's precision and create construct-irrelevant variance (Messick, 1989). Content validity is especially critical for criterion-referenced tests, such as educational assessments, where test scores often guide decisions about individual qualifications or competencies. Lack of adequate content validity may lead to gaps in assessment coverage, meaning the test could fail to capture essential aspects of the construct, undermining the interpretability of the results and any decisions based on them (Downing & Haladyna, 2004).

The interest in this aspect of validity, which is indeed essential for ensuring that psychometric tests are both representative and comprehensive in relation to their intended constructs, has grown in recent year (Spoto et al., 2023; El-Den et al., 2020).

We think that LLMs application in this domain could mark a significant leap in the assessment of content validity, enabling the identification of redundancies or gaps in test coverage through purely algorithmic means. By integrating LLM-based methodologies into psychometric test design, researchers can augment traditional validation approaches with scalable, replicable, and objective tools that enhance the rigor of psychological evaluations. We propose to integrate such methods with more traditional ones, as the expert judgement hinted at before, to provide more robust indications. However, it is essential to explore both the strengths and limitations of these models in comparison to human experts, as this represents the critical focus of our work. While language models offer the potential for scalability, objectivity, and replicability, human experts bring contextual understanding, nuanced judgment, and domain-specific knowledge that algorithms may lack. By evaluating the capabilities of language models against expert judgments, this study aims to identify where artificial intelligence excels, where it falls short, and how the two approaches can complement one another to enhance the assessment of content validity. Content validity is often assessed by judges, and we believe that this process can be complemented by artificial intelligence tools. In other words, content validity, typically reliant on expert judgment, involves evaluating whether a test adequately covers all dimensions of the construct it aims to measure; along with the human experts we want to investigate if modern artificial intelligence technique, particularly methods developed for the analysis of texts and language, can enhance and help in the assessment of the process of content validation.

Due to the great number of tests and items freely available, and to the possible impact for a general audience, the domain of personality assessment provides a suitable benchmark to introduce new approaches. Personality research has been widely used as a testing ground for the development of increasingly sophisticated language models (Hussain et al. 2024; Abdurahman et al., 2023; Nilsson et al., 2024; Giannini et al., 2024). Recent efforts have focused on using language models to predict human responses to personality items and to analyze the factorial structure and taxonomy of personality tests (Wulff & Mata, 2024; Milano et al., 2025). Notably, two language models specifically developed for personality research have recently been introduced, achieving high performance in predicting item-scale relationships and correlations between items (Wulff & Mata, 2024; Hommel & Arslan, 2024).

Given the capabilities of language models to semantically represent human language and their increasingly pervasive use in psychological research, this study explores their application in the context of content validation assessment for psychometric tests.

More in detail, in our study we investigate how language models can enhance the processing and content validation of personality tests using two well-established instruments: the Big Five Questionnaire (BFQ) (Caprara et al., 2013) and the Big Five Inventory (BFI) (John et al., 1991). These tests were chosen because they are among the most widely used tools for measuring personality and capture distinct aspects of test construction through linguistic approaches. The BFQ focuses on describing behaviors and tendencies through language, while the BFI is designed to reveal nuances in personality using concise lexical descriptors.

Starting from these two tests, we analyze the differences in content understanding between language models and humans. First, a group of graduate psychology students served as expert validators, rating each test item according to its alignment with the factors being measured, using the content validity ratio (CVR) as a metric. Using these expert ratings as a baseline, we evaluated the performance of state-of-the-art language models in assessing content validity compared to human experts. Because our experts are Italian, we utilized the Italian versions of the two tests for the expert validation process. Subsequently, we tested the language models using both the Italian and English versions of the tests to evaluate their performance across different languages.

Additionally, we propose a novel language model specifically designed for personality assessment, which utilizes the International Personality Item Pool (IPIP) during the model training. We compare the performance of our model against existing models developed for personality research purposes.

We analyze in depth how these models work, confronting with humans and showing weak and stronger points in the contest of content validity. We observe that, if correctly applied, these models have the potential to reduce the dependency on subjective evaluations, enhance reproducibility, and offer a new procedure to be adopted to streamline the content validation process in psychometric test development.

## 2. Methods

In this section, we outline the methods used to evaluate content validity through both human experts and semantic analysis models. We begin by providing a detailed description of the tests and the procedures employed to assess content validity using expert evaluations. Following this, we introduce the language models used in the study and describe the process of developing a language model specifically tailored for personality assessment. Finally, we present the methods used to automatically assess content validity of questionnaire items using these language models.

### 2.1. Content Validity of the Big Five Questionnaire and Big Five Inventory

To evaluate content validity using both human validators and AI-based methods, we selected two well-established and validated personality assessments: the Big Five Questionnaire (BFQ) and the Big Five Inventory (BFI).

The BFQ (Big Five Questionnaire) is a psychometric tool developed by Caprara, Barbaranelli, and Borgogni (1993) to measure personality traits according to the Five-Factor Model. It assesses Extraversion, Agreeableness, Conscientiousness, Emotional Stability (Neuroticism), and Openness to Experience, each divided into two facets for a detailed profile. Unlike other personality measures like the BFI or NEO-PI, the BFQ emphasizes behavioral descriptions rather than abstract lexical traits, focusing on observable actions and tendencies.

The questionnaire consists of 132 items rated on a 5-point Likert scale. However, due to its length, we randomly selected 50 of the 132 items, with 10 items for each construct, to maintain participant engagement.

The BFQ is particularly valued for its focus on behaviors, which distinguishes it from other trait-focused tools; we chose BFQ for this peculiar property that differentiate it from other personality test and permit us to analyze the level of validity of this kind of items in human and artificial validators.

The BFI (Big Five Inventory) is a 44-item questionnaire designed to measure the Big Five personality traits—Extraversion, Agreeableness, Conscientiousness, Neuroticism, and Openness (John et al.

1991). It uses short, lexical descriptors (e.g., "I see myself as someone who is talkative"), making it concise and easy to administer. The BFI is widely used in research for its brevity and general applicability, though it lacks the detailed subdimensions provided by instruments like the NEO-PI. The main difference between the BFI and BFQ lies in their focus: the BFI uses lexical terms to define personality traits, while the BFQ emphasizes behavioral descriptions of actions and tendencies. This makes the BFQ more aligned with practical and observable behaviors, whereas the BFI provides a more abstract, language-based assessment of traits. Additionally, the BFQ includes subdimensions for each trait, offering greater specificity compared to the BFI's general approach. The complete list of the items used is reported in the supplementary material.

For both questionnaires, human experts rated each item based on its alignment with the five constructs. The level of agreement among the experts was then evaluated using the Content Validity Ratio (CVR) framework.

### 2.2. Content validity ratio

The Content Validity Ratio (CVR) is a statistic used in psychometrics to quantify how much agreement there is among experts regarding the relevance of test items to hypothesized underlying constructs (Lawshe, 1975). The CVR is commonly used to assess the content validity of tests, questionnaires, or surveys to ensure that the items measure what they are intended to measure. It is especially useful in early stages of test development to refine or select items that are valid representations of the construct being measured. A panel of subject matter experts, also called validators, reviews each item on the test or questionnaire. For each item, they determine whether it is *essential*, *useful but not essential*, or *not necessary* for assessing the target construct. For each item, the Content Validity Ratio is calculated based on the proportion of experts who consider the item to be essential, formally:

$$CVR = \frac{(n_e - N/2)}{N/2}$$

Where $n_e$ is the number of experts who rated the item as essential, and N is the total number of experts. The CVR range from -1 to 1.

For an item to be deemed valid, its Content Validity Ratio must meet or exceed a minimum threshold, which varies based on the number of experts involved. Lawshe (1975) provided a table of critical values for different panel sizes, and according to Lynn (1986), using more than 10 experts is likely unnecessary. In our study, we used the CVR calculated from a panel of 10 experts—graduate students in Psychology at the University of Naples Federico II (60% female and 40% male)—as the baseline for an automated content validity procedure. This procedure was developed using deep learning models to analyze sentences, specifically the test items in our case.

### 2.3. Computational embeddings

In Natural Language Processing (NLP), embeddings represent words, phrases, or even entire sentences as dense vectors of numbers in a high-dimensional space. These vectors are designed to capture the meaning, context, and relationships between words or sentences in a way that machine learning models can understand and process (Bengio et al., 2003; Srivastava et al., 2018). The values in these vectors are not arbitrary but are carefully structured to encode semantic and contextual information from text. By training embeddings on large text datasets, they can encapsulate rich information about language, enabling algorithms to convert symbolic forms (like words) into sub-symbolic representations (vectors) for effective processing.

Traditional word embeddings, such as Word2Vec (Mikolov et al., 2013), have a key limitation: they assign a single static vector to each word, regardless of its context. This means that the same embedding is used for a word, even if its meaning varies depending on the sentence. However, with the advent of transformer-based models (Vaswani et al., 2017), contextual embeddings have become the standard. Models like BERT (Bidirectional Encoder Representations from Transformers) (Devlin et al., 2019) and GPT (Generative Pretrained Transformers) (Conneau et al., 2017) produce embeddings that dynamically adjust based on the context in which a word appears, allowing them to generate semantically meaningful representations of both words and sentences (Akhtar et al., 2019; Liu et al., 2019).

In this work, we used embeddings derived from Sentence-BERT (SBERT) (Reimers and Gurevych, 2019), a class of BERT-based models specifically trained to calculate the semantic similarity between sentences. BERT is pretrained on large corpora, including the English Wikipedia and the Google Books dataset, using two main tasks: predicting randomly masked words in a sentence and predicting whether two sentences follow each other. After pretraining, BERT can be fine-tuned with additional output layers to handle specific tasks.

SBERT extends BERT by optimizing it for generating semantically meaningful sentence embeddings. It employs Siamese Neural Networks, a structure that trains on pairs or triplets of sentences, adjusting the embeddings to position semantically similar sentences closer together in the embedding space while pushing dissimilar sentences further apart. Over time, several improved versions of the base SBERT model have been developed, with benchmarks comparing their performance in tasks like sentence similarity (Reimers and Gurevych, 2020).

In our study, we utilized a transformer-based BERT model: Microsoft's MPNet (Masked and Permuted Pre-trained Network) and its variant, Multilingual MPNet (Song et al. 2020). MPNet, a member of the BERT family, is designed to generate high-quality embeddings for various NLP tasks, including sentence similarity, classification, and information retrieval. It is trained on large-scale datasets consisting of diverse sources, providing a general-purpose understanding of language. The multilingual version of MPNet, however, is trained on smaller datasets tailored to multiple languages. Since our validators are Italian native speakers and the tests we used were in Italian, we initially employed the multilingual version of MPNet. To investigate potential differences, we translated the items into English and tested them with the English version of MPNet. We anticipated that the English model would perform better, as it benefits from significantly larger training datasets compared to the multilingual version. This discrepancy is a known issue (Li et al., 2024; Wu and Dredze 2020), as English-language models are often trained on far more extensive corpora than those available for other languages. To address this, we conducted experiments with both the multilingual and English MPNet models.

Furthermore, our ultimate goal is to develop a specialized personality model capable of capturing the semantic similarity between personality-related items. This domain-specific task differs from the general-purpose focus of standard language models trained on diverse sources. To achieve this, we fine-tuned the English MPNet model specifically for understanding personality assessment items, leveraging the availability of large-scale English datasets for training.

In the next section, we review previous attempts to construct fine-tuned models for personality assessment and introduce our approach.

### 2.4. Creating the Personality Fine-Tuned Model

The first attempts to fine-tune language models for predicting relationships among personality items has been carried by Hommel and Arslan, 2024; and Wulff and Mata, 2024.

Hommel and Arslan (2024), developed the SurveyBot3000 model by fine-tuning a pre-trained MPNet model, employing a multi-stage process to address nuanced semantic relationships, including polarity calibration and domain-specific adaptation. In the first stage, they modified the Stanford Natural Language Inference (SNLI) corpus, adding directionality to cosine similarities for contradictory pairs to better reflect opposing psychological constructs, such as introversion versus extraversion. The second stage involved training on a curated dataset of 90,424 item pairs drawn from publicly available personality inventories, ensuring high-quality representation and rigorous cross-validation by partitioning the data into training, validation, and test sets with no overlap. They further tested the model on a robust holdout dataset of 87,153 item pairs from Bainbridge et al. (2022) to evaluate generalizability. Hommel and Arslan also cleaned item texts to eliminate unrelated linguistic features, such as adverbs of frequency, to ensure the model captured only trait-relevant semantics.

Wulff and Mata (2024) fine-tuned MPNet to adapt it to predicting unsigned correlations between personality items, irrespective of whether they were positively or negatively correlated. Their approach relied on diverse datasets, including four inventories from the Open-Source Psychometrics Project (16PF, BIG5, FFM, and HEXACO), data from the NEO inventory by Kajonius and Johnson, and the Eugene-Springfield Community Sample, covering over 1 million responses in total. They designed a training set of 200,000 examples by pairing items and using absolute Pearson correlations as the target variable. Special care was taken to standardize text formatting across datasets, such as removing the pronoun "I" from item phrasing to ensure uniformity. They balanced the data by oversampling high-correlation pairs to counteract their rarity in the raw datasets, and used the sentence-transformers library to fine-tune MPNet with a CosineSimilarityLoss function.

Our approach is more straightforward, we do not preprocess the data and do not implement polarity-specific adjustments to enhance model sensitivity to reversed-items as Hommel and Arslan, 2024. We do not use human correlation as target, as in Wulff and Mata, 2024; but we relies uniquely on semantic similarity between items.

To create a model as capable as possible of classifying semantically similar items together, we developed a method to fine-tune the Microsoft all-MPNET model on personality items. We used the International Personality Item Pool (IPIP) as the dataset to fine-tune the model.

IPIP is an open-access collection of personality items, scales, and measures used for psychological research and assessment. It was designed to provide researchers and practitioners with a resource to measure various personality traits using standardized tools. The IPIP is especially valuable because it offers a wide range of items based on well-established personality models. Additionally, IPIP is completely open-source, which is why we used this database as the training set to fine-tune the general-purpose model on personality items.

The IPIP database contains over 3,320 items derived from more than 250 scales. Since our goal was to create a model specifically trained to detect similarities between personality items, we generated all possible item pairs from the IPIP dataset, excluding repetitions and identical items (i.e., no self-pairing). Formally:

$$\binom{n}{k} = \frac{n!}{k!\,(n-k)!}$$

Where n is the total number of items (in our case 3250), and k is the number of items for each combination, in this case 2 for a couple. This led to over 15 million possible items pair.

In order to fine-tune the model, we needed to label the item pairs with target similarity scores to train our language model. To achieve this, we used a Cross-encoder neural network. Cross-encoders are a type of architecture primarily used in Natural Language Processing (NLP) tasks such as text pair classification, sentence similarity, and ranking tasks. Cross-encoders jointly encode two input texts to capture interactions between them and output a similarity score for the input sentences. For our Cross-encoder, we used the Microsoft MS-MARCO-MiniLM-L12-v2 model, a general-purpose model trained to predict similarities between natural language sentences in a range between -1 (very dissimilar sentences) and +1 (very similar sentences).

Once the dataset based on IPIP was created, we fine-tuned the base all-MPNET model. Details of the training procedure are provided in the supplementary material. It is important to note that none of the items from the BFI or BFQ are present in the IPIP database, so we did not need to divide the dataset into training and testing sets. In Figure 1 a schematic representation of the fine-tuning method is reported.

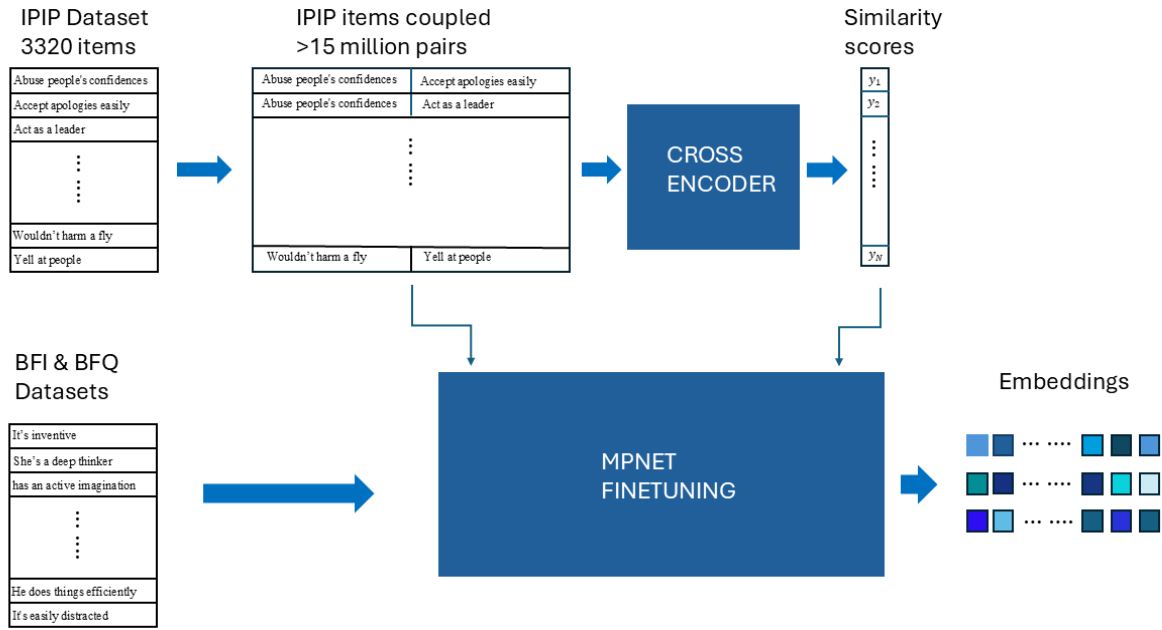

Figure 1. Schematic representation of the fine-tuning method

Our goal is to create a more informative embedding space specifically designed for personality items. We hypothesize that this embedding space will be better at detecting similarities between items and distinguishing between items related to different psychological constructs. To test our hypothesis, we compared the performance of differently fine-tuned MPNet model against general-purpose language models, using experts performance as the baseline accuracy.

In the next subsection, we describe the proposed algorithm for using the language models as validators. This algorithm automatically, and in an unsupervised manner, group semantically similar items under the same latent constructs.

## 2.5. Automated Content Validity method

Our procedure provides the entire list of $N$ items as input to the language model (LLM). The items are then embedded into 768-dimensional vectors, forming a matrix ($N$ x 768), where each row represents the embedded vector of a specific item.

At this stage, we calculate the cosine similarity between each item and all other items, resulting in a $N$ x $N$ square matrix. Unlike human validators, the language model does not directly assign each item to an underlying construct. To address this, we propose a method for automatically classifying each item into a single construct.

We calculate the average cosine similarity (CS) for each item with all items theoretically belonging to each of the five constructs, excluding the item's similarity with itself from the average. To determine the final classification of an item as related to a specific construct, we apply the SOFTMAX function to the average CS across all constructs. The SoftMax function transforms the cosine similarities into probabilities, representing the likelihood that an item belongs to a particular construct. In this way, we select the construct with the highest probability as the representative for each items of the test. We can measure the accuracy of our method by simply dividing the number of items assigned to the right underlying construct for the number of items expected from the theory; doing this way, we can measure how accurate is the prediction for each construct and averaging the accuracy of the constructs retrieve the total accuracy of the model.

## 3. Results

In this section, we analyze how the validators rate each questionnaire item in terms of its alignment with the constructs the test is designed to measure. We then examine how Large Language Models

(LLMs) capture semantic similarities between items. Based on these similarities, we propose a method for assigning each item to a construct and test this method using two LLMs: one multilingual model, which matches the language of the validators, and one English-based model, which processes items translated into English.

The accuracy of both the validators' and LLMs' responses is evaluated against the true associations between items and constructs, as documented in the literature. Additionally, we test fine-tuned models on their ability to classify BFQ and BFI items, comparing their performance to general-purpose LLMs and validators' ratings.

For clarity, we refer to the model fine-tuned by Hommel and Arslan (2024) as SurveyBot MPNet, the model fine-tuned by Wulff and Mata (2024) as Personality MPNet and our fine-tuned model as Cross-Encoder MPNet (Cross Encoder). To streamline the results section we have included figures only for the human experts and the best-performing method in the main text. Detailed graphs for other methods are available in the supplementary material for reference.

### 3.1. Study 1 BFQ personality test

#### 3.1.1. Experts' rating

Each validator was tasked with rating the items according to the Content Validity Ratio (CVR) methodology, using a 3-point scale: 0 for items deemed not useful, 1 for items considered useful but not essential, and 2 for items judged absolutely necessary to assess the construct. Items were assigned to the construct that received the highest number of positive ratings from the validators.

To evaluate the accuracy of these assignments, we compared the validators' item-construct mappings with the ground truth provided by the BFQ. The total accuracy was calculated as the ratio of correctly identified relationships to the total number of predictions. Overall, the validators correctly matched 82% of the items to their corresponding constructs, demonstrating a strong alignment between most items and their intended constructs.

However, a closer examination of the results reveals variability across constructs (Figures 2 and 3). While the relationships for constructs such as agreeableness, conscientiousness, and openness showed high accuracy, extraversion emerged as an exception, with only 50% of the items being correctly matched. This suggests a weaker connection between the items and the intended construct as identified by the validators. Despite this anomaly, the results remain largely consistent with the expected item-construct relationships hypothesized in the test design, reinforcing the validity of the overall approach.

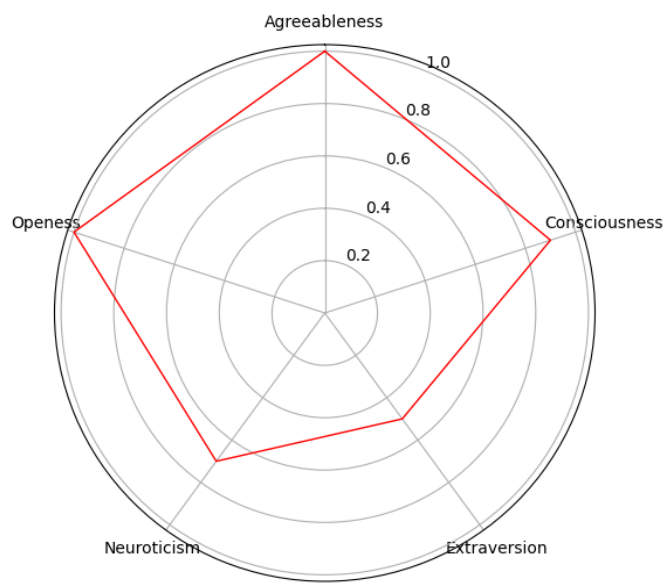

Figure 2. Radar plot of the percentage of the items assigned correctly to the underlying construct from the validators ratings

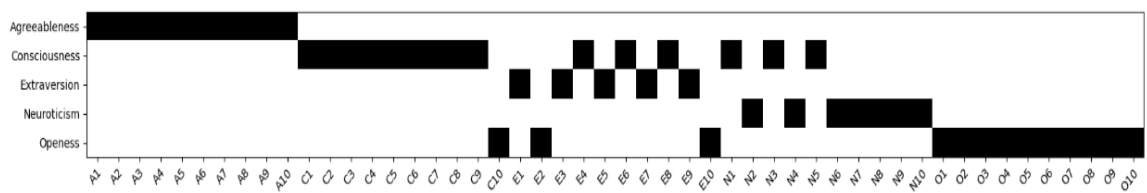

Figure 3. Schematic representation of the items, x-axis, and of the constructs, y-axis. The black square indicates the construct where an item has been assigned from the validators, the items range span from 1 to 10 and the capital letter show the construct to what they belong according to the theory.

### 3.1.2. MPNet Multilingual

A Multilingual Large Language Model was then employed to classify the items according to their underlying constructs. Since the validators are Italian native speakers and the BFQ test is in Italian, we utilized the BERTMultilingual model as our initial approach to derive an automated measure of content validity.

Using this approach, the multilingual model achieved an overall accuracy of 64% in assigning items to the correct constructs. The breakdown of accuracy for each construct is as follows: agreeableness (50%), conscientiousness (70%), extraversion (60%), neuroticism (90%), and openness (50%) (Figures S1 and S2).

Compared to the human validators, this approach demonstrated a notable drop in accuracy, particularly for the constructs of agreeableness and openness, where half of the items were misclassified. These findings highlight the challenges faced by the multilingual model in capturing the nuanced semantic relationships within the constructs when applied to a personality assessment in Italian.

### 3.1.3. MPNet

To evaluate whether the multilingual model was suboptimal, we translated the items into English and repeated the analysis using the MPNet model, which is trained exclusively on English text. Multilingual models often exhibit biases toward specific languages due to the diverse nature of their training datasets and are exposed to smaller amounts of text per language compared to models trained solely in English.

The English-based MPNet model achieved a total accuracy of 70% in classifying the items to their respective constructs. The construct-specific accuracies were: agreeableness (80%), conscientiousness (50%), extraversion (60%), neuroticism (100%), and openness (60%); see Figures S3 and S4.

Compared to the multilingual model, the English-based MPNet showed a slight improvement in overall accuracy, suggesting that training exclusively on English text provides a marginal advantage for this task. This indicates that the multilingual approach may be hindered by the limitations of its training data, particularly when applied to a specialized domain like personality assessment.

### 3.1.4. SurveyBot MPNet

The model developed by Hommel and Arslan (2024) is a fine-tuned version of the base MPNet. To evaluate its performance, we tested this model, referred to as SurveyBot, on the BFQ items to determine whether it outperforms the base MPNet.

SurveyBot achieved a total accuracy of 64% in classifying items to their respective constructs. The construct-specific accuracies were as follows: agreeableness (60%), conscientiousness (30%), extraversion (80%), neuroticism (100%), and openness (50%); Figures S5 and S6.

Compared to the base MPNet model, the fine-tuning resulted in a decrease in overall accuracy for the BFQ test. While the classification accuracy for neuroticism items remained unchanged, the model exhibited a decline in performance for the constructs of conscientiousness, openness, and agreeableness. However, it showed an improvement in classifying extraversion items. These results suggest that the fine-tuning process may have introduced biases or limitations that reduced its effectiveness for some constructs in this specific test

### 3.1.5. Personality MPNet

We tested the model developed by Wulff and Mata (2024), a fine-tuned version of MPNet (referred to as Personality MPNet), to evaluate its content validity accuracy on the BFQ items.

Personality MPNet achieved an overall accuracy of 72% in classifying items to their respective constructs. The construct-specific accuracies were as follows: agreeableness (60%), conscientiousness (80%), extraversion (60%), neuroticism (100%), and openness (60%); see Figures S7 and S8.

Compared to the base MPNet model, Personality MPNet slightly improves overall performance. Notably, it shows a substantial improvement in classifying neuroticism items, with accuracy increasing from 50% to 100%. However, this comes at the cost of a decrease in accuracy for agreeableness items, dropping from 80% to 60%. Performance for the other constructs remains consistent with the base model.

### 3.1.6. Cross Encoder MPNet

Finally, we tested our fine-tuned version of the MPNet model (referred to as Cross Encoder MPNet) on the BFQ items. Cross Encoder MPNet achieved an overall accuracy of 80% in classifying items to their respective constructs. The construct-specific accuracies were: agreeableness (80%), conscientiousness (70%), extraversion (70%), neuroticism (100%), and openness (80%); (Figures 4 and 5).

This represents a significant 10% improvement over the base MPNet model, with enhanced performance across all constructs. The classification accuracy achieved by Cross Encoder MPNet is notably close to the performance of the human validators, demonstrating the potential effectiveness of this fine-tuning approach in improving content validity assessments

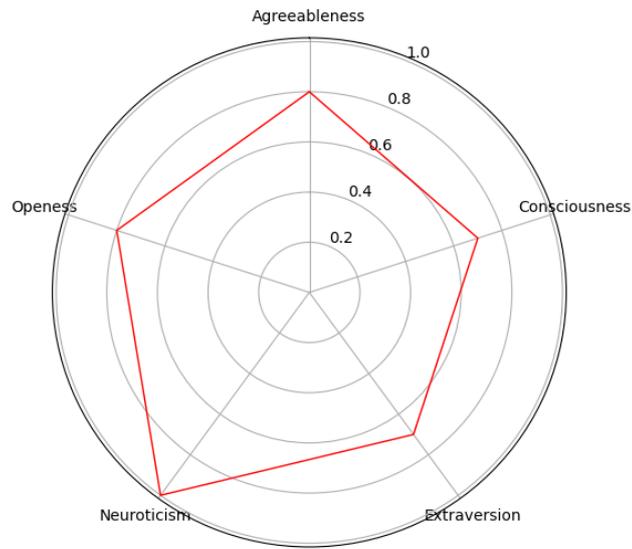

Figure 4. Radar plot of the percentage of the items assigned correctly to the underlying construct from the validators ratings

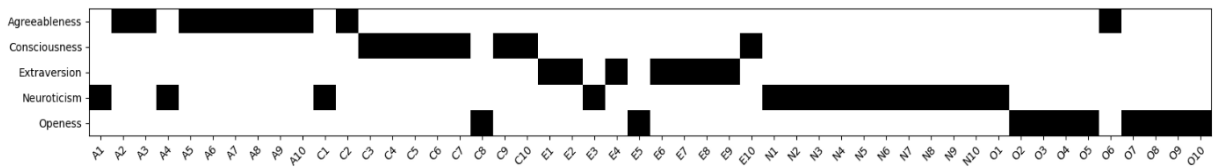

Figure 5. Schematic representation of the items, x-axis, and of the constructs, y-axis. The black square indicates the construct where an item has been assigned from the validators, the items range span from 1 to 10 and the capital letter show the construct to what they belong according to the theory.

### 3.2. Study 2 BFI personality test

#### 3.2.1. Experts' ratings

We replicated the methods outlined in Study 1 and used the item-construct assignments from the BFI as ground truth to measure how accurately the validators identified the correct relationships. The total accuracy was calculated as the ratio of correct predictions to total predictions.

The validators correctly matched 72% of the items with their respective constructs. The construct-specific accuracies were as follows: agreeableness (62.5%), conscientiousness (100%), extraversion (62.5%), neuroticism (50%), and openness (87.5%); (Figures 6 and 7).

While the results align reasonably well with the expected item-construct relationships for openness and conscientiousness, the accuracy is notably lower for the other constructs. This suggests that the validators were more successful in matching items to these two constructs, but there were challenges in accurately associating items with the remaining constructs.

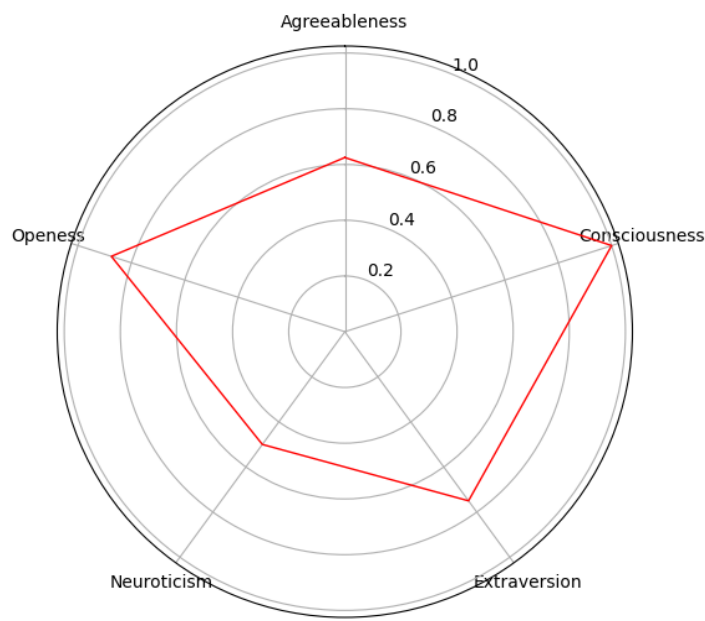

Figure 6. Radar plot of the percentage of the items assigned correctly to the underlying construct from the validators ratings

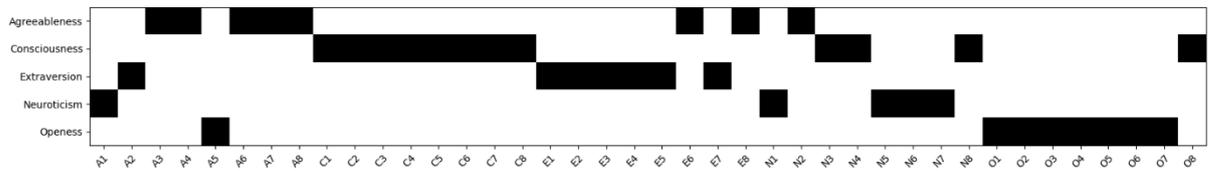

Figure 7. Schematic representation of the items, x-axis, and of the constructs, y-axis. The black square indicates the construct where an item has been assigned from the validators, the items range span from 1 to 10 and the capital letter show the construct to what they belong according to the theory.

### 3.2.2. MPNet Multilingual

For the BFI items, the MPNet Multilingual identified the item-construct relationships with a total accuracy of 77.5%. The construct-specific accuracies were as follows: agreeableness (75%), conscientiousness (75%), extraversion (62.5%), neuroticism (50%), and openness (100%); see Figures S9 and S10.

Interestingly, unlike the results with the BFQ, the multilingual model outperformed the validators in this study, providing a more precise mapping of the items to their corresponding constructs. This suggests that the multilingual model may be more effective in handling the BFI items, potentially due to differences in the structure and wording of the two tests.

### 3.2.3. MPNet

As with the BFQ test, the BFI items were translated into English, and the MPNet model was applied to retrieve the embeddings.

Using this model, we achieved a total accuracy of 80%, with the following construct-specific accuracies: agreeableness (87.5%), conscientiousness (75%), extraversion (87.5%), neuroticism (50%), and openness (100%); Figures S11 and S12.

Once again, the English-language model outperformed the multilingual model, demonstrating that English-based language models tend to have a greater capacity for analyzing language. This suggests that the larger and more specialized training datasets available for English language models contribute to their superior performance in this context.

### 3.2.4. SurveyBot MPNet

SurveyBot, when applied to the BFI, achieved an accuracy of 75%, with construct-specific accuracies as follows: agreeableness (87.5%), conscientiousness (62.5%), extraversion (87.5%), neuroticism (62.5%), and openness (100%); see Figures S13 and S14.

Once again, this model performed worse than the base MPNet on the BFI items, failing to improve item classification within the given constructs. It is likely that the fine-tuning procedure has been detrimental to this type of analysis, possibly introducing biases or limitations that hindered the model's effectiveness for the BFI test.

### 3.2.5. Personality MPNet

The Personality model, when applied to the BFI, achieved an impressive accuracy of 97.5%, with construct-specific accuracies as follows: agreeableness (100%), conscientiousness (87.5%), extraversion (100%), neuroticism (100%), and openness (100%); (Figures 7 and 8).

This represents a significant improvement over the base MPNet model used for fine-tuning. Nearly all items were correctly classified, highlighting the effectiveness of the approach employed by Wulff and Mata (2024). Their fine-tuning strategy appears to be particularly well-suited for lexical-based items, such as those found in the BFI setup.

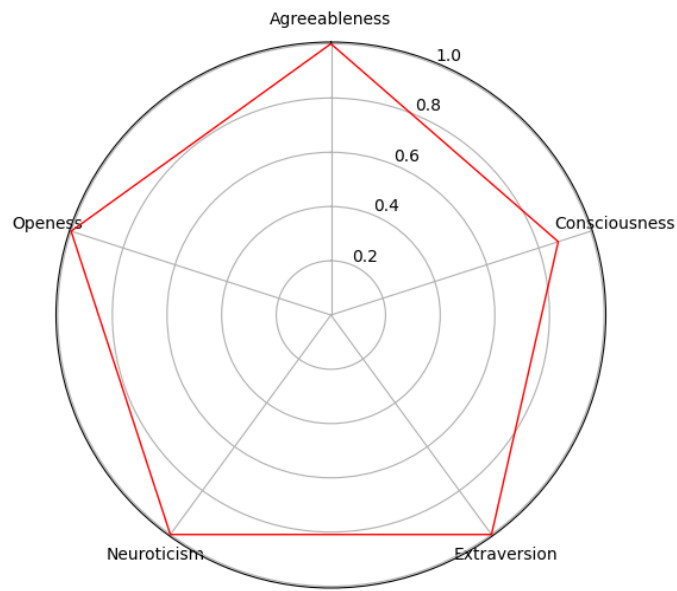

Figure 7. Radar plot of the percentage of the items assigned correctly to the underlying construct from the Personality MPNet model

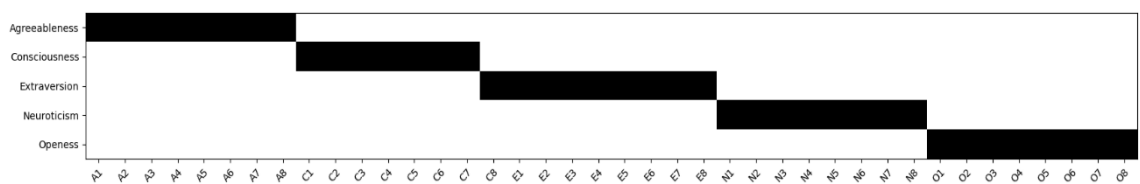

Figure 8. Schematic representation of the items, x-axis, and of the constructs, y-axis. The black square indicates the construct where an item has been assigned from the Personality MPNet model, the items

range span from 1 to 10 and the capital letter show the construct to what they belong according to the theory.

### 3.2.6. Cross Encoder MPNet

Our Cross Encoder model, based on MPNet, achieved a total accuracy of 82.5% in classifying the items to their related constructs. The construct-specific accuracies were as follows: agreeableness (75%), conscientiousness (75%), extraversion (62.5%), neuroticism (100%), and openness (100%); see Figures S15 and S16.

In the BFI setup, our model outperformed the base MPNet, though it did not reach the performance level of the Personality MPNet model, which proved to be the most effective at handling the BFI items.

## 4. Discussion

| BFQ | agreeableness | conscientiousness | extraversion | neuroticism | openess | Total |
|---|---|---|---|---|---|---|
| humans | 100% | 90% | 50% | 80% | 100% | 84% |
| MPNet-ML | 50% | 70% | 60% | 90% | 50% | 64% |
| MPNet | 80% | 50% | 60% | 100% | 60% | 70% |
| SurveyBot | 60% | 30% | 80% | 100% | 50% | 64% |
| Personality | 60% | 80% | 60% | 100% | 60% | 72% |
| Cross Encoder | 80% | 70% | 70% | 100% | 80% | 80% |

| BFI | agreeableness | conscientiousness | extraversion | neuroticism | openess | Total |
|---|---|---|---|---|---|---|
| humans | 62.5% | 100% | 62.5% | 50% | 87.5% | 72% |
| MPNet-ML | 75% | 75% | 62.5% | 50% | 100% | 77.5% |
| MPNet | 87.5% | 75% | 87.5% | 50% | 100% | 80% |
| SurveyBot | 87.5% | 62.5% | 87.5% | 62.5% | 100% | 75% |
| Personality | 100% | 87.5% | 100% | 100% | 100% | 97.5% |
| Cross Encoder | 75% | 75% | 62.5% | 100% | 100% | 82.5% |

Table 1 and 2 summarize the results reported in previous section, providing the accuracy for each construct and the total accuracy; each model is reported along with the results of human experts. The results of this study illuminate the complex interplay between human expertise and artificial intelligence in evaluating content validity. By comparing human validators to various models of Large Language Models (LLMs), we reveal how the characteristics of the test items affect the accuracy and utility of these approaches. A notable distinction in this study lies in the nature of the personality test items. The BFI items predominantly represent lexical terms, serving as samples of the linguistic domains associated with personality constructs. In contrast, the BFQ items are behavioral in nature, describing specific actions or tendencies that exemplify the constructs they measure. This difference had significant implications for the performance of both human validators and LLMs.

Human validators perform better in the BFQ test, achieving an 82% accuracy rate. This performance reflects their strength in interpreting the nuanced behavioral cues embedded in the BFQ items. Behavioral descriptions often require contextual and experiential understanding, which humans naturally possess, enabling them to discern subtle alignments between items and constructs. However, the human performance dropped to 72% for the BFI, likely because humans may struggle to consistently identify the implicit lexical patterns that define construct-related terms without additional theoretical support.

In contrast, LLMs demonstrated greater proficiency with the BFI's lexical items. The models' embeddings, particularly when fine-tuned or trained on extensive text corpora, are adept at capturing the semantic relationships among words. For instance, the baseline English MPNet achieved 80% accuracy with the BFI, outperforming its performance on the BFQ (70%). Similarly, the Personality MPNet, fine-tuned using lexical personality data, achieved an impressive 97.5% accuracy on the BFI, underscoring its effectiveness in identifying lexical relationships. However, LLMs underperforms with the behavioral complexity of the BFQ items, where in comparison each models score less than in the BFQ, reflecting the challenges in bridging the gap between linguistic embeddings and behavioral semantics.

The distinction between lexical and behavioral domains also influenced construct-specific performance. Constructs such as neuroticism and openness were consistently well-classified across both tests, likely because their definitions—whether lexical or behavioral—are conceptually coherent and semantically distinct, even not fine-tuned models has optimal or close-optimal accuracy in classifying those constructs. Conversely, constructs like extraversion and agreeableness posed challenges, particularly in the BFQ, where the behavioral nature of the items may introduce overlapping or ambiguous cues that are difficult for both humans and models to disentangle. For the BFI, the lexical clarity of these constructs likely contributed to the models' improved accuracy, particularly with fine-tuned approaches.

The differential performance across tests and methods has important implications for psychometric test development. The results reinforce the importance of human expertise in evaluating items that reflect complex, contextualized behaviors. While LLMs can process text with high semantic precision, the interpretation of behaviors requires a broader cognitive framework that encompasses personal experience and cultural knowledge. Even if some approaches like the Cross Encoder barely reach human performance, LLMs appear to struggle in evaluating behavioral-based items.

The superior performance of LLMs on lexical items highlights their potential as a complementary tool in content validation, particularly for tests like the BFI that draw on linguistic domains. Fine-tuned models like Personality MPNet, which specialize in lexical constructs, can serve as reliable, scalable alternatives or supplements to human judgments in these contexts.

The challenges faced by LLMs with the BFQ highlight a gap in current natural language processing techniques. Future research should explore embedding strategies that better capture the semantic and contextual nuances of behavioral descriptions. For instance, fine-tuning on datasets explicitly focused on behavioral language may enhance the models' capacity to interpret such items. Furthermore, the effects of the length of the items on the LLMs performance is a variable that needs to be analyzed, the BFQ items are sensibly longer than the BFI items and this can have impact when the embeddings are retrieved by the language models.

While the study highlights key insights, certain limitations must be acknowledged. First, the reliance on ground truths from literature assumes the correctness of these item-construct mappings, which may not fully account for construct evolution or cultural variability. Second, translating BFQ items into English for some LLMs may have introduced linguistic distortions that affect their interpretability of the original behavioral content.

## 5. Conclusion

Due to the advancement of deep learning in Natural language processing, LLMs are becoming pervasive and widely used in psychological assessment. Here we tried to provides insights into how these models can be useful for content validity in psychometrics, analyzing how state of the art models, and their specialization explicitly tailored for personality research, perform against a human baseline. We compare fine-tuned models on the top of the same base model but with different training approach; furthermore, we analyze how the nature of test items—lexical or behavioral—interacts with the capabilities of human validators and AI-based methods in assessing content validity. By systematically comparing human performance to that of Large Language Models (LLMs) across two distinct personality tests, the BFQ and BFI, we highlighted strengths and limitations inherent to each approach.

Our work shows how different LLMs asses content validity from psychological items compared to humans. We use two different versions of the Big-Five personality, BFI and BFQ, each specifically designed with items more lexical or behavioral related to the underlying construct, to compare fine-tuning strategies applied to a common base model, MPNet, with the performance validation of human validators to assess content validity. This allowed us to directly evaluate how specific training to the base architecture impact the model's ability to predict content validity. The results reveal a clear hierarchy in the effectiveness of fine-tuning approaches. For instance, the Personality MPNet, fine-tuned using a pipeline tailored for lexical relationships, achieved outstanding accuracy on the BFI items (97.5%), far surpassing both the base MPNet and other fine-tuned versions, such as SurveyBot and our Cross Encoder MPNet. Conversely, while our model demonstrated significant improvements in accuracy for the BFQ, it highlights the challenges faced by LLMs in interpreting behavioral items. The variation in model performance underscores the importance of tailoring fine-tuning methods to the specific demands of the task. The Personality model's success can be attributed to its domain-specific training on lexical constructs, which allows it to discern semantic relationships with high precision. In contrast, SurveyBot's less consistent results suggest that fine-tuning procedures must be carefully aligned with the end task, as misalignment can reduce the generalizability of the model. These findings point to the need for further research into fine-tuning methods, particularly for tasks involving complex, context-dependent items such as those in the BFQ.

Human validators, by contrast, excelled in assessing the behavioral items of the BFQ, achieving an accuracy of 82%—a result that models have yet to match. This demonstrates the critical role of human expertise in interpreting the nuanced, contextual nature of behavioral constructs. However, their performance was less robust for the lexical items in the BFI, where they achieved 72% accuracy, suggesting that lexical relationships are more consistently captured by computational methods. These complementary strengths highlight the potential of hybrid approaches, combining the contextual understanding of human experts with the scalability and precision of fine-tuned LLMs.

These findings have important implications for psychometric test development. First, the success of specific fine-tuning approaches demonstrates that AI models can be highly effective in tasks requiring semantic precision, provided that the training strategy is appropriately designed. This is particularly relevant for tests that rely on lexical constructs, where AI-based validation can serve as a scalable, reliable complement to human judgments. Second, the persistent challenges faced by models in evaluating behavioral items suggest a need for further innovation in this area, including the development of embeddings specifically tailored for behavioral language.

Future research and applications should build on these insights to refine and expand the use of AI in psychometric validation. Developing task-specific fine-tuning pipelines and hybrid validation systems can optimize accuracy and efficiency while maintaining the interpretative depth offered by human expertise. By addressing these challenges, we can harness the strengths of both humans and AI to advance the scientific rigor and utility of psychometric assessments.